 \documentclass{emulateapj}

\newcommand{\mpc}{\, {\rm Mpc}}
\newcommand{\kpc}{\, {\rm kpc}}
\newcommand{\pc}{\, {\rm pc}}
\newcommand{\kmps}{\, {\rm km \, s^{-1}}}

\newcommand{\Myr}{\,{\rm Myr}}
\slugcomment{{{\sc Accepted to ApJ:} November 30, 2012}}

\shorttitle{Geometric offsets across spiral arms in M51}
\shortauthors{Melissa Louie}

\begin{document}

\title{Geometric offsets across spiral arms in M51: Nature of gas and star formation tracers}
\author{Melissa Louie and Jin Koda}\affil{Department of Physics and Astronomy, Stony Brook University, Stony Brook, NY 11794-3800}
\email{melissa.louie@stonybrook.edu}

\author{Fumi Egusa}\affil{Department of Space Astronomy and Astrophysics, Institute of Space and Astronautical Science, Japan Aerospace Exploration Agency, Japan}

\begin{abstract}
We report measurements of geometric offsets between gas spiral arms and associated
star forming regions in the grand-design spiral galaxy M51.
These offsets are a suggested measure of the star formation timescale after the compression
of gas at spiral arm entry.
A surprising discrepancy, by an order of magnitude, has been reported in recent offset
measurements in nearby spiral galaxies. 
Measurements using CO and H$\alpha$ emission find large and ordered offsets in M51. On the contrary, small or non-ordered offsets have been found using the HI 21cm and
$24\micron$ emissions, possible evidence against gas flow through spiral arms,
and thus against the conventional density-wave theory with a stationary spiral pattern.
The goal of this paper is to understand the cause of this discrepancy.
We investigate potential causes by repeating those previous
measurements using equivalent data, methods, and parameters.
We find offsets consistent with the previous measurements
and conclude that the difference of gas tracers, i.e., HI versus CO, is the primary cause.
The HI emission is contaminated significantly by the gas photo-dissociated
by recently-formed stars and does not necessarily trace the compressed gas,
the precursor of star formation.
The HI gas and star forming regions coincide spatially and tend to show small offsets.
We find mostly positive offsets with substantial scatter between CO and H$\alpha$,
suggesting gas flow through spiral arms (i.e., density-wave) though the spiral pattern may not necessarily be stationary.

\end{abstract}


\section{Introduction}

The density wave theory \citep{linshu64} has been a central paradigm for the formation of spiral structures in galaxies. \citet{rob69} suggested that star formation (SF) in spiral galaxies is triggered by a spiral density wave and predicted that H$\alpha$ emission around newly-formed stars should be offset from and appear downstream of the gas spiral arm due to flow through the density wave. The offsets would not only be evidence of the density wave, but would allow measurement of the angular speed of the spiral pattern and the timescale for star formation \citep{egu04}, given a velocity difference between the matter and the spiral pattern and assuming that the SF timescale is constant. The offset and circular velocity can be measured observationally as a function of radius, and should obey the relationship,
\begin{equation}
\Delta\Theta(r)_{A \rightarrow B} = (\Omega(r)-\Omega_{p})\Delta t
\end{equation}
where $\Delta\Theta(r)$ is the azimuthal offset angle between gas spiral arms (A) and star forming regions (B), $\Omega(r)$ is the angular velocity of the gas, $\Omega_{p}$ is the angular pattern speed, and $\Delta t$ is the time that the gas takes to evolve into young massive stars after spiral arm entry.

A discrepancy has emerged in recent studies of offsets \citep{tam08, egu09, foy11}.
 In an analysis of $\sim 10$ galaxies, \citet[][hereafter E09]{egu09} derived offsets larger by an order of magnitude in general than \citet[][T08]{tam08}.
More surprisingly, \citet[][F11]{foy11} reported primarily non-ordered offsets compared to those predicted by the standard density-wave theory  (Eq. 1).
This discrepancy directly affects the estimates for $\Omega_{p}$ and $\Delta t$, and thereby their physical implications for star formation.
Indeed, the measured star formation timescales are inconsistent: $5 - 30 \Myr$ for E09 versus $1 - 4 \Myr$ for T08. 
The shorter timescale may imply very short lifetimes for giant molecular clouds (GMCs), i.e., their exceedingly rapid formation and destruction and associated rapid star formation (T08), while the longer timescale may be consistent with the gravitational collapse of gas at typical giant molecular cloud gas densities (E09).

This study explores the potential causes of the discrepancy. Two major differences among these studies are the choices of emission tracers and the methods of offset measurement. E09, and originally \citet{egu04}, used CO (J=1-0) and H$\alpha$ as tracers of the dense gas and recently-formed young stars, respectively.
T08 and F11 used HI 21 cm emission and IR 24$\micron$ emission to trace the locations of the compressed gas and young massive stars, respectively. T08 and F11 also compared low-resolution CO and 24$\micron$ emission on a subset of galaxies. F11 also investigated other tracers such as UV emission to trace SF and 3.6 $\micron$ emission to trace the old stellar population.
CO is the established tracer of the dense star forming molecular gas, while HI, if concentrated around spiral arms, may coincide with the dense gas.
H$\alpha$ observations typically provide a higher angular resolution than observations at 24$\micron$.  However, H$\alpha$ observations could miss dust-obscured star forming regions, whereas 24$\micron$ emission suffers very little from extinction.

The methods of offset measurement could also be the cause of the discrepancy. \citet{egu04} and E09 compared the location of emission peaks by eye. This ``Peak Tracing'' Method is intuitive and easy to apply, but is possibly biased. T08 and F11 used an automated cross-correlation method to avoid any human bias. This ``Cross-Correlation'' Method is apparently less subjective, but surprisingly its application leads to inconsistent results even when the same data are used (T08 and F11).

The offset measurements may constrain emerging theories of spiral arm formation. In $N$-body simulations, the spiral arms appear to form as material arms instead of density waves \citep{wada11}. Spiral structures are constantly formed and broken apart by gravitational shearing  \citep{goldreich65}. In this case, no clear offset would be expected across the spiral arm. The stellar and gas spiral arms are most likely co-spatial, the gas being pulled toward local stellar potential minima, instead of passing through them as predicted by the density-wave theory \citep{wada11}.
\citet{dobb10} performed hydrodynamic simulations in stationary and transient stellar spiral structures (i.e., density-wave spiral versus material spiral, respectively) and found that the offset depends strongly on the nature of the stellar spiral structures.
The ordered offsets measured by \citet{egu04}, T08, and E09 support the density wave theory in its simplest form with a single pattern speed, while the result of F11 could indicate a transient spiral structure without a fixed pattern speed. Resolving the discrepancies in the offset measurements is therefore increasingly important.

In this paper, we reexamine the measurements of the geometric offsets by T08, E09, and F11, using the previously-adopted four tracers, HI, CO, 24$\micron$ and H$\alpha$ emission, and using both the Peak Tracing and Cross-Correlation methods. We focus on M51 to clarify the primary causes of the discrepant results from previous studies. We conclude that the discrepancy comes from the different tracers,
mainly because the HI 21 cm emission traces the gas dissociated by young stars rather than the parental gas for star formation.
We will demonstrate that the two measurement methods provide consistent results if appropriately applied. In addition, we will discuss some caveats in applying the offset model for deriving the pattern speed and the SF timescale.

M51 is among the galaxies analyzed by T08, E09 and F11. However, they did not find consistent geometrical offsets $\Delta \Theta$ (and, almost equivalently, the star formation timescale $\Delta t$). For example, T08 measured $\Delta t = 3.4 \pm 0.8 \Myr$ ($\Omega_{p}= 21 \pm 4 \kmps \kpc^{-1}$), while E09 obtained $\Delta t = 13.8 \pm 0.7 \Myr$ ($\Omega_{p}= 40 \pm 4 \kmps \kpc^{-1}$), i.e. a factor of 4 difference in $\Delta t$ (a factor of 2 in $\Omega_{p}$).
A summary of the tracers used and results can be found in Table \ref{tab:previouswork}.

The data for the gas and star formation tracers are described in $\S 2$. The offset measurement methods of the previous studies are discussed in $\S 3$.
Our results with four emission tracers, HI, CO, $24\micron$, and H$\alpha$, are discussed in $\S 4$.
The measurement methods and tracers are compared in $\S 4.2$ and $\S 4.3$, respectively.
The cause of the discrepancies among previous measurements 
and the limitations of the model are discussed in $\S 5$.
A summary of this work is given in $\S 6$.

\section{Data}

In order to elucidate the causes of the discrepancy, we compare the tracers (i.e., HI, CO, 24$\micron$ and H$\alpha$) previously used in T08, E09 and F11. The HI data are from the HI Nearby Galaxy Survey \citep[THINGS]{walter08}, an HI survey made with the Very Large Array at the National Radio Astronomy Observatory. We use the data reduced with natural weighting. The CO data are from \citet{koda09,koda11} and were obtained by combining Combined Array for Research in Millimeter Astronomy (CARMA) and Nobeyama 45m telescope observations as a part of the CARMA-Nobeyama nearby galaxies (CANON) CO (J=1-0) survey.
The 24$\micron$ and H$\alpha$ data are from the Spitzer Infrared Nearby Galaxies Survey \citep[SINGS]{kenn03}. The 24$\micron$ observations were made with the Spitzer Space Telescope and the H$\alpha$ observations were taken with the 2.1m telescope at The Kitt Peak National Observatory. The spatial resolutions of HI, CO, 24$\micron$ and H$\alpha$ data are 
$\simeq$6'' ($280 \pc$ at 9.6 Mpc), 4'' ($160 \pc$), 5.7''($256 \pc$), and 1.9'' ($88 \pc$), respectively.

The HI and 24$\micron$ data are the same as used in T08 and F11. The CO data have a higher resolution and sensitivity than the data from the BIMA Survey of Nearby Galaxies \citep[BIMA-SONG; ][]{hel03} that were used in T08 and E09; however, both resolve the molecular spiral arms in M51, and there should not be much difference in the offset analysis. F11 used much lower resolution (13'') CO (J=2-1) data from the HERA CO-Line Extragalactic Survey \citep[HERACLES;][]{leroy09} in their analysis.
The H$\alpha$ image is not the same as the one that E09 used, though it has a similar resolution. We do not expect a significant difference between the two, because the analysis is weighed significantly to high signal-to-noise regions in bright spiral arms. A slight improvement in the S/N does not change the positions of the bright peaks.
We adopt the following parameters for M51: a distance ($D$) of $9.6 \mpc$, a major axis position angle (P.A) of $22\degr$ and an inclination angle ($i$) of $20\degr$ \citep{sofue99}. We will test the different sets of these parameters and confirm that the choice does not affect our conclusions.

\section{Offset Measurement and Analysis}\label{sec:method}
The procedure for the offset method has two major steps: (1) measurements of azimuthal offsets between gas and star formation tracers, and (2) the determination of a star formation timescale and pattern speed using the measured offsets and a rotation curve. Our primary focus is to investigate possible causes of the discrepancy in step (1).
The discrepancy between the previous offset measurements (i.e., large offsets in E09, small offsets in T08, and no ordered offsets in F11) could be due to differences in the methods employed and/or in the adopted tracers of dense gas and star forming regions.
We briefly summarize the two methods here.

\begin{figure}
\includegraphics[scale=.75]{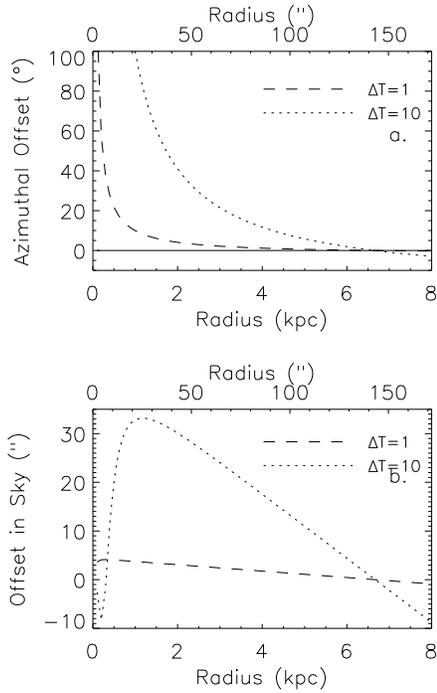}
\caption{
The expected offsets determined from  Eq. (1) for a constant $v_{gas}$
of $200 \kmps$ and $\Omega_{p}$ of $30 \kmps \kpc^{-1}$.
Two star formation timescales are assumed, 
$\Delta t = 1 \Myr$ and $10 \Myr$.
({\it a}) The azimuthal offset in the galactic disk plane as a function of radius.
({\it b}) The angular offset in the sky.
The offset is expected to be largest at smaller radii and approaches zero
with increasing radius.
\label{fig:expectedoffset}}
\end{figure}

Both methods are based on theoretical predictions of the density-wave theory \citep{rob69}.
Assuming a flat galactic rotation curve and a constant pattern speed, the relative velocity between the gas/stars and the spiral pattern changes with radius. If the timescale of SF after spiral arm compression is a constant, this differential motion results in a spatial offset between the gas spiral arm and star forming regions, and the offset changes with radius. If we measure the offset as a function of radius, the SF timescale and pattern speed can be determined \citep{egu04}. 
Figure \ref{fig:expectedoffset} shows two model plots of the expected offset versus galactic radius, assuming SF timescales of 1 and 10 Myr and other parameters appropriate for M51 (i.e., pattern speed and rotation velocity). Figure \ref{fig:expectedoffset} (a) shows the azimuthal offset in the galactic disk, while Figure \ref{fig:expectedoffset} (b) shows the offset in the sky. The azimuthal offset is expected to be large at small radii and approximately zero at large radii.

For both methods, the four images, HI, CO, 24$\micron$ and H$\alpha$, are regrided to the same pixel size using the Multichannel Image Reconstruction, Image Analysis, and Display (MIRIAD) software package \citep{sau95} and are deprojected using 
the MIRIAD task ``deproject''. 
 The deprojected images were transformed into phase diagrams (radius, azimuth) using the MIRIAD task ``polargrid''.
After the production of the phase diagrams, the analysis continues differently for each method as discussed below. 

We wrote our own IDL programs to reproduce the different measurement techniques of E09, T08, and F11, following the detailed procedures discussed in these papers.

\subsection{Peak Tracing Method}\label{sec:peaktracing}
\begin{figure*}
\includegraphics[scale=0.5]{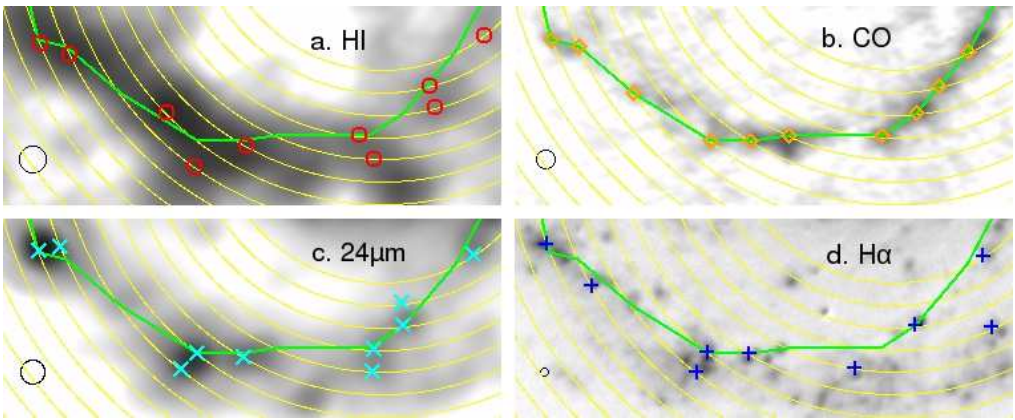}
\caption{
Images of a portion of Arm 1 in (a) HI, (b) CO, (c) 24$\micron$, and (d) H$\alpha$.
The positions of peaks identified by the PT method in radial bins (centered on the concentric arcs)
are plotted on the images. 
The black circles in the lower-left corners indicate the resolutions of the data.
The gas flow direction is counter-clockwise. As a reference, the CO peaks are connected with green lines in all panels.}
\label{fig:spiralarm}
\end{figure*}

A measurement method for the geometrical offset was developed by \citet{egu04} and updated in E09. The offset is measured by eye as the azimuthal angular offset between the emission peaks of gas and star forming region tracers. This {\it Peak Tracing method} starts by binning the phase diagrams in the radial direction with an appropriate bin size to match the spatial resolution of all tracer images (here we use a bin size of 5''). The emission peaks are visually determined in an azimuthal intensity profile of each bin.
The offsets are defined as the azimuthal shifts of the peaks between the two tracers. We find that this visual identification tends to pick up only peaks with a high signal-to-noise ratio. 
Obviously, some radii are not included in the analysis if no emission peaks could be identified in their azimuthal profiles.
Figure \ref{fig:spiralarm} shows examples of identified peaks in all the four tracers in a part of Arm 1: (a) HI , (b) CO, (c) 24$\micron$, and (d) H$\alpha$. Gas is flowing in a counter-clockwise direction.

\subsection{Cross-Correlation Method}\label{sec:crosscorrelation}

\begin{figure}
\includegraphics[scale=1]{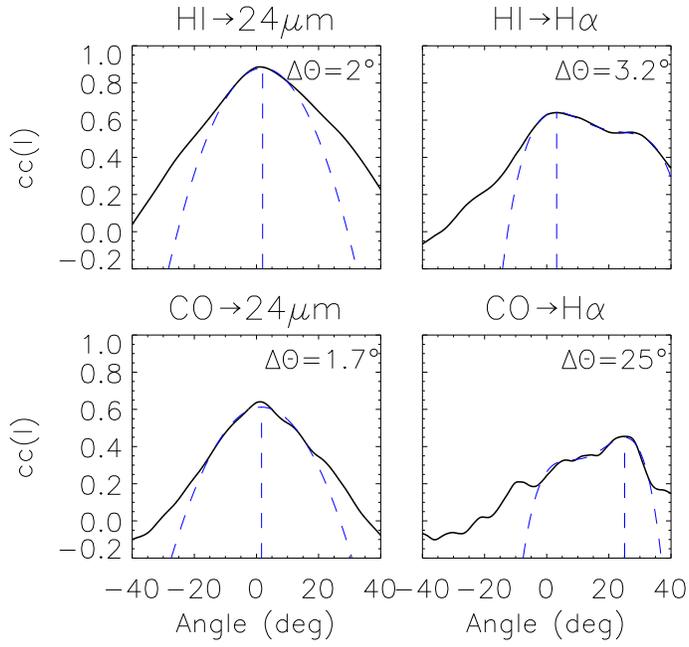}
\caption{Examples of the cross-correlation function for the four combinations of the tracers at a radius of 2.51 kpc. The dashed curve show a forth degree polynomial fit around the maximum peak. The dashed blue line shows the location of the maximum peak. The offset value is listed in the top right of each panel for each tracer combination.}
\label{fig:crosscore}
\end{figure}

T08 cross-correlates the azimuthal profiles of the gas and SF tracers in each radial bin. They define a normalized cross-correlated function, 
\begin{equation}
cc_{x,y}(l)=\frac{\Sigma_{k}[(x_{k}-\bar{x})(y_{k-l}-\bar{y})]}{\sqrt{\Sigma_{k}(x_{k}-\bar{x})^{2}\Sigma_{k}(y_{k}-\bar{y})^{2}}},
\end{equation}
where $x_{k}$ and $y_{k}$ are the fluxes in gas and SF tracers (in T08 and F11, HI and 24 $\micron$, respectively).
$\bar{x}$ and $\bar{y}$ are the mean values of the fluxes. $l$ is the lag (or the shift) in pixel units between the two profiles. 
The denominator normalizes the cross-correlation function based on the variance of each data set; this normalized cross-correlation function is useful when adopting a cutoff value to remove low signal-to-noise data at different wavelengths. 
In principle, the cross-correlation function shows a maximum at the value of $l=l_{\rm max}$ that best aligns the two profiles -- finding the peak of the cross-correlation function should be enough to obtain $l_{\rm max}$. In practice, we fit a polynomial to the profile of the cross-correlation function around the peak to reduce the effect of noise.
T08 uses a fourth-degree polynomial, and we follow their method. Examples of the cross-correlation function for a radial bin (at the galactic radius of 2.51 kpc) are shown in Figure \ref{fig:crosscore} for each set of tracer combinations.
The offset is simply $\Delta\Theta(r) = l_{max}(r)$.
T08 adopted a lower threshold (i.e., 0.2) for the normalized cross-correlation function to reject measurements with low significance and F11 used a threshold of 0.3.  We set the threshold for an acceptable cross-correlation value to 0.3 to avoid low signal-to-noise measurements.

\subsection{Parameters and Errors} \label{sec:errors}

We apply both the Peak Tracing Method and the Cross-Correlation Method to two gas tracers (CO and HI) and two SF tracers (H$\alpha$ and 24$\micron$). The phase diagrams are binned radially in 5" increments for all data.  This bin size is chosen to match the resolutions of the HI and 24$\micron$ data.
We denote the spatial offset between the gas (tracer $A$) and star forming regions ($B$) as $\Delta \Theta_{A\rightarrow B}$ and measure $\Delta \Theta (r)_{HI \rightarrow 24\micron}$, $\Delta \Theta(r)_{HI \rightarrow H\alpha}$, $\Delta \Theta(r)_{CO \rightarrow 24\micron}$, and $\Delta \Theta(r)_{CO \rightarrow H\alpha}$.
If multiple  peaks exist in the cross correlation function, the largest is chosen, since by definition this is the location where the peaks of the spiral arms are best aligned.
For the Peak Tracing Method, the locations of any multiple peaks are checked against maps of M51 to make sure that they correspond to the spiral arms instead of emission in the inter arm regions \citep{koda09}.

Two types of error could be introduced in the analyses: (1) due to uncertainties in determination of peak positions in each tracer, and (2) due to the misassociation of the peaks in gas and SF (e.g., in the case of multiple peaks).
The first type of error is easy to estimate and is simply related to the resolution of the images in both methods.
This relation may not be so obvious in the case of the Cross-Correlation Method, but, for example, if there are point sources convolved with point-spread functions (PSF; i.e., resolutions) the cross-correlation function is simply a multiplication of the two PSFs with a variable lag. Its width (or the associated error) is therefore determined by the sizes of the resolutions. In all the data. we measure only the positions of bright peaks and the positional accuracy should be a small fraction of the resolutions. We here conservatively adopt half of the resolution as the error. This choice provides an error similar to that estimated by E09, but much larger than that of T08.
The second type of error is difficult to estimate, and we neglect it for now. This error, however, should appear as scatter in our measurements.

\section{Results}\label{sec:results}

We qualitatively compare the images of the different tracers before making quantitative measurements. To distinguish the two arms, we define Arm 1 as the spiral arm that is further from the companion galaxy and Arm 2 as the one that connects to the companion.

Figure \ref{fig:phase} shows phase diagrams of the four tracers. Contours of star formation tracers (H$\alpha$ and 24$\mu$m) are shown on gas tracer images (CO and HI). Gas flows in the direction of increasing azimuthal angle.
Figure \ref{fig:spiralarm} shows the relative locations of the peaks in the four tracers with respect to CO (i.e., green line).
If we compare the peak locations of the SF tracers, H$\alpha$ and 24$\micron$,
with respect to CO peaks, the H$\alpha$ emission appears downstream of the CO emission
when an offset is seen.
This is consistent with the findings in E09.
The 24$\micron$ peaks show a similar trend and appear to be leading the CO peaks. 
Most important are the locations of the HI peaks as they are mostly at the leading side of CO, the well-established tracer of star-forming gas. The HI peaks appear to be spatially closer to the peaks of the two star formation tracers.
This trend is inconsistent with the hypothesis in T08 and F11
that the HI traces the compressed, parental gas for SF.
The H$\alpha$ and 24$\micron$ peaks appear at similar locations in most radial bins,
although occasionally they show offsets.
The offsets between CO and H$\alpha$/24$\micron$ are evident over a range of radii, while HI shows little or no offsets from H$\alpha$ and 24$\micron$ emission. Another notable point is that CO is predominantly in the inner part of the galaxy, while  HI is in the outer part.

We note that the spiral arms do not form narrow smooth lines;
on closer inspection they are fragmented and wiggle back and forth
between the leading and trailing sides \citep[most obviously seen in CO; see also][]{koda09}.
This causes an additional intrinsic error in the offset measurements.

We apply the Peak Tracing Method and Cross-Correlation Method to these four data sets,
and quantitatively verify the cause of the discrepancies in the previous measurements. 
We omit the very central part (radii $\lesssim 2$ kpc), since the spiral structure is tightly wound and we cannot reliably associate the star-forming gas and resultant young stars in that region.

\subsection{Comparisons to Previous Works}\label{sec:previous_work}
Figure \ref{fig:discrepancy} compares the offsets measured using the same tracers and methods as the previous studies. The left column shows the results from the Peak Tracing method applied to one spiral arm (i.e., Arm 1; as in E09). The right column shows the same offsets but from the Cross-Correlation Method applied to both arms (as done in T08 and F11). The offsets between HI and 24$\micron$ are shown in the top panels, and the offsets between CO and H$\alpha$ are shown in the bottom panels.

The measured offsets in Figure \ref{fig:discrepancy} are consistent with the previous measurements of T08, E09, and F11 in terms of the ranges of their amplitudes. E09 compared CO and H$\alpha$ data using the Peak Tracing Method and found offsets up to 25-30 $\degr$ in Arm 1. Figure \ref{fig:discrepancy} ({\it bottom left}) shows our corresponding measurement. The amplitude range of our offsets between CO and H$\alpha$ is similar to that in E09 (see their Figure 5, Arm 1). 
Our range in Arm 2 is also consistant with that of E09 (see Figure \ref{fig:offsets}). E09 found irregular offsets between CO and H$\alpha$ along Arm 2, and our measurements for Arm 2 see similar irregularities, which are discussed later in this section.
T08 analyzed HI and 24$\micron$ using the Cross-Correlation Method (see their Figure 4) and found smaller offsets (mostly less than 5-10 $\degr$), except for the innermost radii (some large offsets of 15-20$\degr$ $\lesssim 2$ kpc). 
In our comparison of HI and 24$\micron$ (Figure \ref{fig:discrepancy}, {\it top right}) we find similar small offsets, most of which are under 10 $\degr$ in $\gtrsim 2$ kpc, consistent with T08 results. 
We were unable to measure offsets at the innermost radii (discussed above).
F11 also measured offsets between HI and 24$\micron$ using the Cross-Correlation Method and found that they are mostly under $10 \degr$ (their Figure 10), which is comparable to our amplitude range. 
The amplitude of the offsets found by T08, E09, and F11 agrees with our measured offsets when we use the same set of tracers and measurements method.
Our measurements reproduce the amplitude discrepancy in the previous studies.

Figure \ref{fig:offsets} shows plots of all of the azimuthal offsets measured for each set of tracers as a function of galactic radius for the Peak Tracing method ({\it left two columns}) and the Cross-Correlation method ({\it middle two columns}). We measure the offsets in the two arms separately in these four columns. T08 and F11 analyzed the two arms together to measure one offset at each radius. For comparison, we also analyze the two arms together and show the plots in Figure \ref{fig:offsets} ({\it right column}). E09 found a strange trend in Arm 2, where the offset increases with radius (as opposed to the prediction in Figure \ref{fig:expectedoffset}). Our results for Arm 2 from the Peak Tracing method show similar behavior. Along Arm 2, the emission from the SF tracers appear to be scattered over a large area (though mostly at the leading side of the arm). Therefore, we discuss only Arm 1 in the following analysis when we analyze the two arms separately. 

T08 also analyzed CO and 24$\micron$ data (their Figure 7) and found that the offsets are even smaller than those seen between HI and 24$\micron$. We found the contrary. The offsets between CO and 24$\micron$ in Figure \ref{fig:offsets} ({\it third row}) are larger compared to those between the HI and 24$\micron$ ({\it top row}). F11 also compared a lower-resolution CO image with a 24$\micron$ image and noted possible evidence for ordered offsets in M51, though they did not quantify the CO$\rightarrow$24$\micron$ offsets further.
Indeed, compared to their Figure 11,
we find even larger offsets at the small radii (Figure \ref{fig:discrepancy} ({\it top right})).
Even if we smooth our CO data to a 13"  resolution to match F11's,
we still find larger offsets than F11.
F11 also analyzed the two arms separately, but report no difference.
Again, this is inconsistent with our results. Part of the reason for this difference may be that they used higher transition CO (J=2-1) data while we used CO (J=1-0). The higher transition emission might be excited due to heating by young stars \citep{koda12}. If that is the case, CO (2-1) should appear closer to SF tracers (i.e. smaller offsets).

\subsection{Method Comparison}\label{sec:result_method}

The two measurement methods provide, in general, consistent results and do not appear to be the cause of the discrepancy.
Figure \ref{fig:offsetoffset} directly compares the measured offsets in Arm 1 from the Peak Tracing Method and the Cross-Correlation Method (i.e., the points in Figure \ref{fig:offsets}) and shows good agreement, with some outliers.
The dotted lines have a slope of one. Two solid lines in each panel show $\pm 5 \degr$ from the dotted line. We find that 70\% of the data points are within $\pm 5 \degr$.
Except for some outliers, most of the measured offsets are comparable, and neither method appears to bias the offsets in any systematic way in any of the panels. The two methods give roughly consistent offsets. The top left panel only has 3 out of 19 radii where the two measurement methods do not show a correlation. For the top right panel 6 out of 19 points do not agree. The bottom left panel has 5 points out of 19 and the bottom right panel has 6 out of 17 points that do not show a correlation.

E09 discussed that the two arms show quantitatively different offsets and analyzed them separately with the Peak Tracing Method. The difference between the two arms is clear in Figure \ref{fig:offsets} ({\it first and second columns}). Since the two methods generally provide consistent results when applied to single arms, it is natural to expect differences in the measured offsets when the two arms are analyzed simultaneously using the Cross-Correlation Method. We see this in our analyses of both arms simultaneously (Figure \ref{fig:offsets} ({\it fifth column}) and separately (Figure \ref{fig:offsets} ({\it third and forth columns})) with the Cross-Correlation Method. We note that both T08 and F11 analyzed the arms simultaneously, while F11 also isolated individual arms; the difference between the two arms was not mentioned in either case. From our work, regardless of the measurement method (the Peak Tracing or Cross-Correlation Methods) and the analysis (the two arms separately or simultaneously), the offsets are mostly positive, but do not appear ordered.

\subsection{Tracer Comparison}\label{sec:result_tracers}

The choice of tracers appears to be the dominant cause of the discrepancy. Large differences are seen among the panels in Figure \ref{fig:offsets} ({\it left column}) and \ref{fig:offsets} ({\it middle column}), where the two arms are analyzed separately. Most remarkably, HI and $24\micron$ -- the combination employed by T08 and F11 show the smallest, if non-zero offsets for M51. Similarly, very small and zero offsets are seen between HI $\rightarrow$ H$\alpha$. If we use CO as a tracer of the dense, compressed gas (e.g., CO $\rightarrow$ H$\alpha$ used by \citet{egu04} and E09), the offsets are much larger. Clearly, the difference in gas tracer is the dominant cause.

Figure \ref{fig:tracercompare} shows a direct comparison of the offsets between gas tracers and SF tracers in a given radial bin. The left column shows the offsets from the Cross-Correlation Method, while the right column shows those from the Peak Tracing method. The top panels compare the offsets found using HI with H$\alpha$ and CO with H$\alpha$. The bottom panels show the same, but for 24$\mu$m. 
The offsets between HI and the SF tracers are mainly zero; however at the same radii the offsets between CO and the SF tracers are larger. In addition, when non-zero offsets are measured for both gas tracers, the offsets between CO and the SF tracers appear larger than the offsets between HI and the SF tracers. This is evident in Figure \ref{fig:tracercompare} as most data points appear in the top-left parts of the panels. Occasional outliers exist, but there are no more than two in each panel. 

\begin{figure*}
\includegraphics[scale=1]{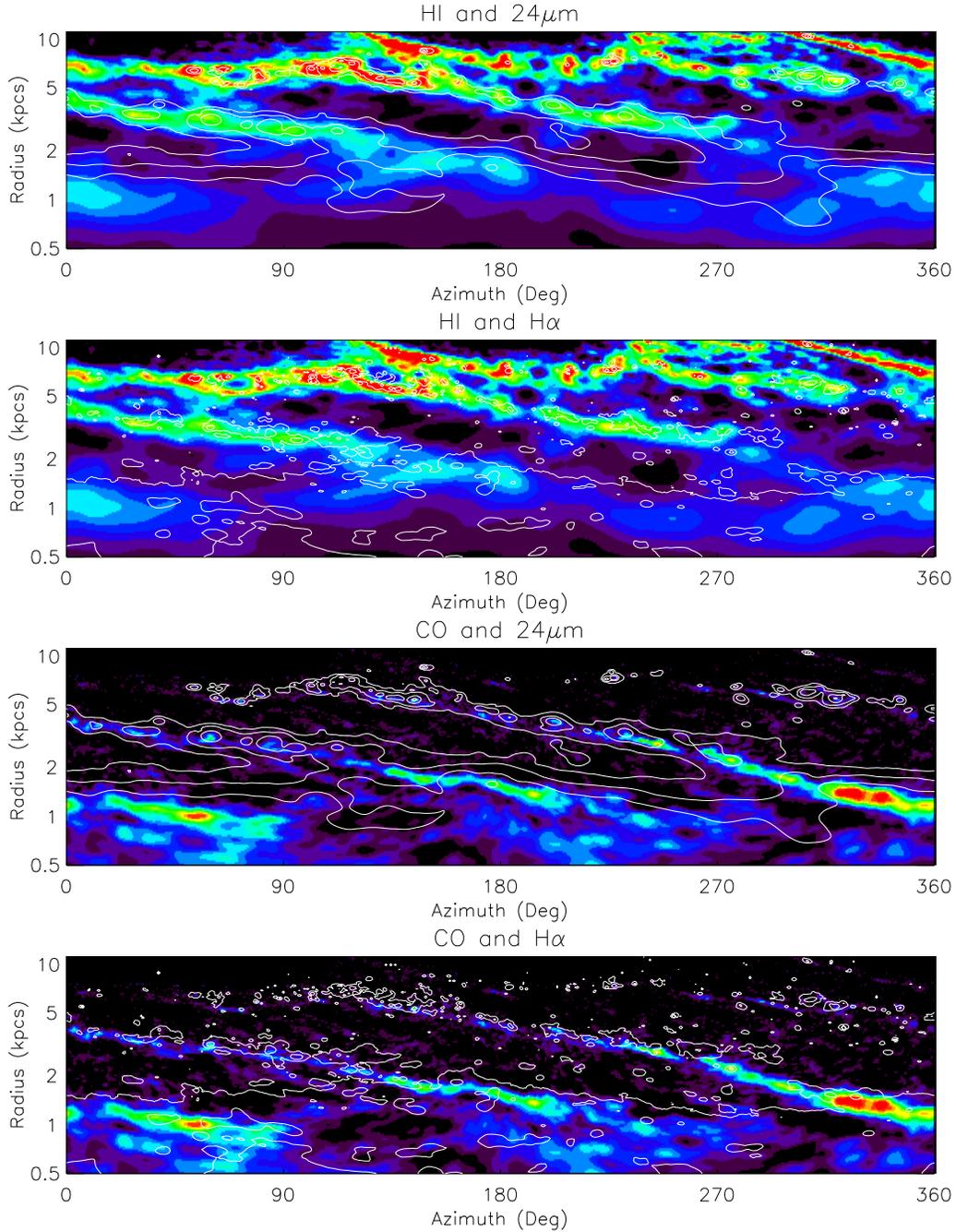}
\caption{Phase diagrams of combinations of the four tracers for comparison of our results with those in T08, Fl1 ({\it top-right}) and E09 ({\it bottom-left}). The offsets are measured using the methods adopted in T08, F11 ({\it top panels}) and E09 ({\it bottom panels}).}
\label{fig:phase}
\end{figure*}

\begin{figure}
\includegraphics[scale=.6]{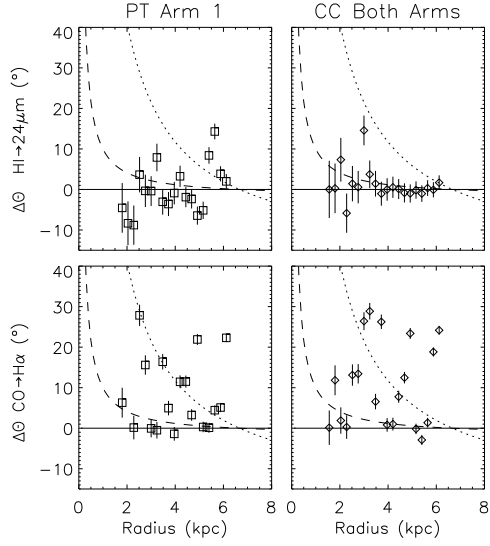}
\caption{Azimuthal offsets between gas and SF tracers. The top row shows the offsets between HI and 24$\micron$ emission  ($\Delta \Omega_{\rm HI \rightarrow 24\micron}$), while the bottom row shows those between CO and H$\alpha$ ($\Delta \Omega_{\rm CO \rightarrow H\alpha}$).
The Peak-Tracing (PT) method is applied to Arm 1 for the left column, and the Cross-Correlation (CC) method is applied to both spiral arms simultaneously
for the right column. The top-right panel is directly comparable to the measurements by T08 and F11 (as it uses the same tracers and method).
The bottom-left panel is comparable to the measurement by E09.
}
\label{fig:discrepancy}
\end{figure}

\begin{figure*}
\includegraphics[scale=1]{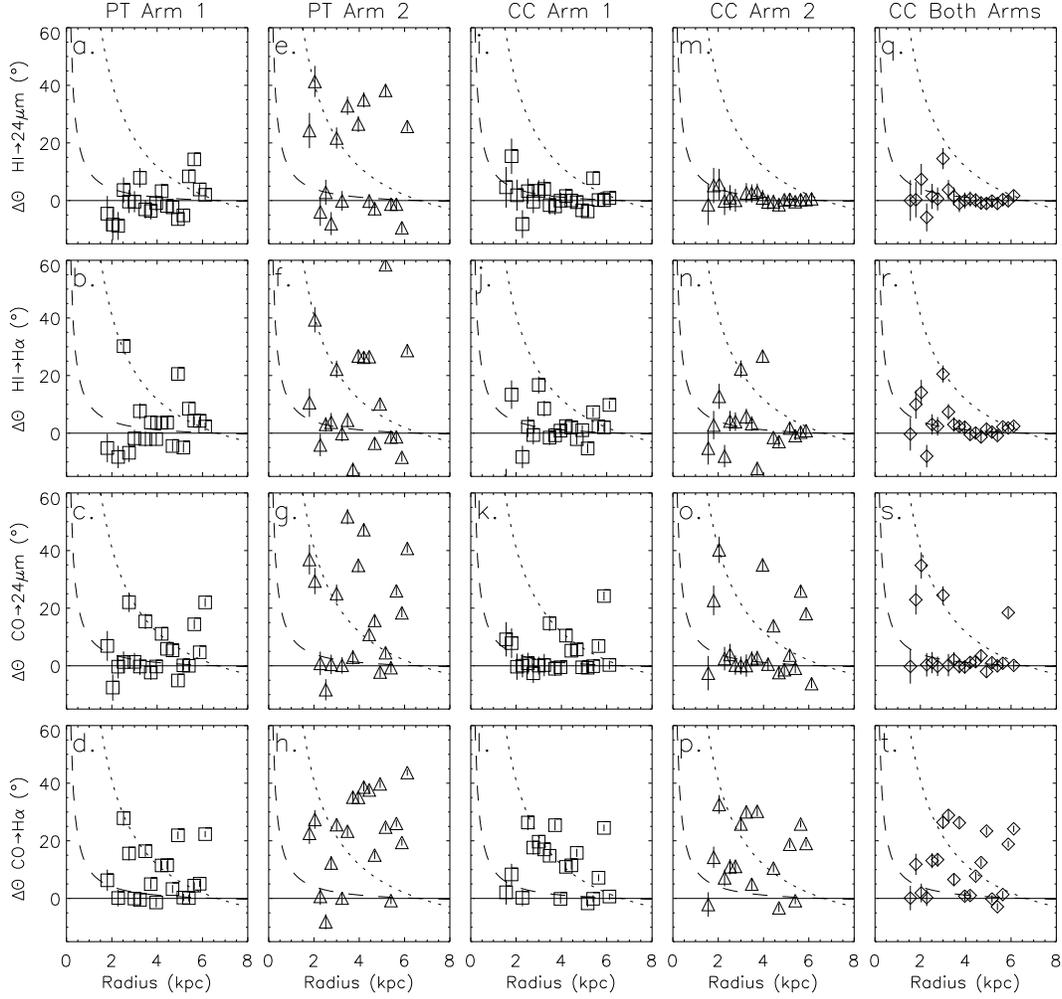}
\caption{Azimuthal offsets between gas and SF tracers as a function of radius.
The two spiral arms are analyzed both separately and simultaneously with
the Peak Tracing (PT) method and Cross-Correlation (CC) method.
{\it Left two columns:} the offsets determined by the PT method with the two arms analyzed separately,
{\it Middle two columns:} by the CC method with the two analyzed separately, and
{\it Right column:} by the CC Method with the two analyzed simultaneously.
The rows from top to bottom show the offsets between
HI $\rightarrow$ 24$\micron$,
HI $\rightarrow$ H$\alpha$,
CO $\rightarrow$ 24$\micron$, and
CO $\rightarrow$ H$\alpha$.
The model predictions of offsets for star formation timescales of 1 and 10 Myr are overplotted (see Figure \ref{fig:expectedoffset}).
\label{fig:offsets}}
\end{figure*}

\begin{figure}
\includegraphics[scale=.75]{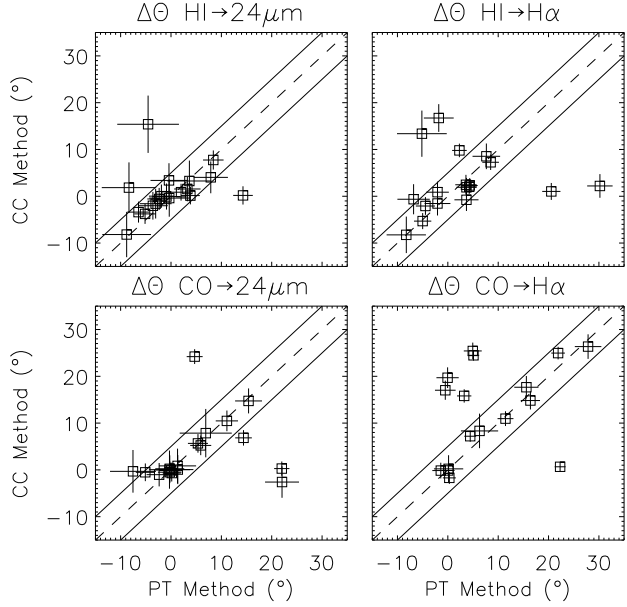}
\caption{
Comparisons between the offsets using the Cross-Correlation (CC) and the Peak Tracing (PT) methods.
Offsets are measured at each radial bin for Arm 1, (see Figure \ref{fig:offsets}).
For all four tracers, the offsets measured with the two methods are in general agreement, with some outliers. Neither method appears to be biased in any systematic way. The fraction of data that show correlation is $\sim 70 \%$. For reference, a dashed line with a slope of one is plotted and two solid lines are plotted to show $\pm 5 \degr$.
\label{fig:offsetoffset}}
\end{figure}

\begin{figure}
\includegraphics[scale=.75]{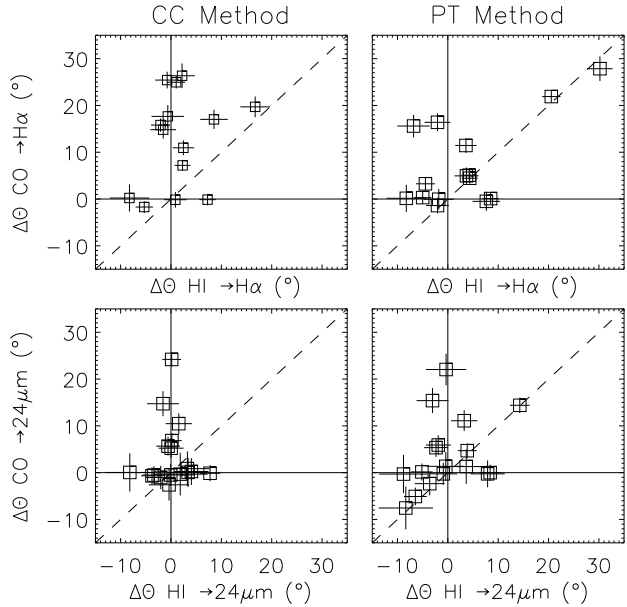}
\caption{Comparisons between the offsets measured using different gas tracers on Arm 1. 
The top row shows the comparisons between $\Delta \Omega_{\rm CO \rightarrow H\alpha}$ and $\Delta \Omega_{\rm HI \rightarrow H\alpha}$,
while the bottom row shows those between $\Delta \Omega_{\rm CO \rightarrow 24\micron}$ and $\Delta \Omega_{\rm HI \rightarrow 24\micron}$.
The CC method is applied in the left column, and the PT method is applied in the right column.
For reference, the dashed line has a slope of one (i.e., the two offsets are equal) and the two solid lines show zero (no offsets).
Most points appear in the top-left of each panel, indicating that the offsets between CO and SF tracers are larger than those between HI and SF.
The offsets between HI and SF tracers are often zero.
}
\label{fig:tracercompare}
\end{figure}

\begin{figure}
\includegraphics[scale=.75]{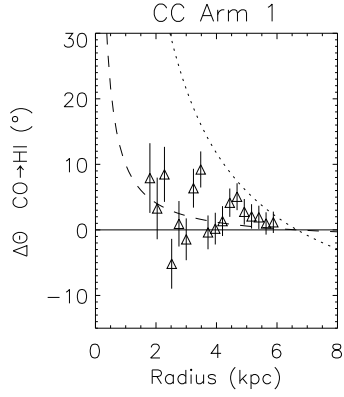}
\caption{Offsets between CO and HI as a function of radius. 
Most offsets are positive, indicating that the HI peaks are on the downstream side of the CO.}
\label{fig:cohicompare}
\end{figure}

There are multiple components that could be contributing to the HI emission, which would give different offsets (and possibly non zero offsets) depending on which component is dominant at the radius in question. For example, HI may trace some of the gas compressed upon entry into a spiral arm, resulting in positive offsets, but would also trace the gas photo-dissociated by newly-formed young stars, leading to zero offsets. 
There is also an extended background component in the HI emission that is virtually everywhere across the galactic disk. We can also find some spatial displacements between HI and CO peaks in Figure \ref{fig:phase} - HI is mostly downstream of CO, indicating that the HI emission is not simply from the dense gas, but multiple sources. This trend is also seen in Figure \ref{fig:cohicompare}, where $\Delta\Theta_{\rm CO \rightarrow HI}$ is plotted against radius. The positive offsets suggest that HI is downstream of the CO. We will discuss this more in \S \ref{sec:cause}.

As for SF tracers, 24$\micron$ and H$\alpha$ peaks are mostly consistent; therefore, the offsets with both tracers are more or less the same in Figure \ref{fig:offsets}. However, there are some noticeable differences at some radii. For example, in Figure \ref{fig:phase} there are radii at which H$\alpha$ peaks appear downstream of $24\micron$ peaks. This is perhaps due to the fact that the distribution of $24\micron$ emission is affected by the distributions of both dense gas/dust and illuminating stars. If newly-formed stars escape from the dense parental gas (as we see in Figure \ref{fig:phase}), the $24\micron$ emission may not trace the locations of the young stars any more, because the peak of dust density is left behind. The relevance of the tracers to the offset analysis is discussed in \S \ref{sec:cause}.

\subsection{Miscellaneous Differences}

\subsubsection{Resolution}\label{sec:resolution}

The offset measurement could be sensitive to the resolution of the data. From Figure \ref{fig:expectedoffset}, we can determine that the azimuthal offset is expected to be  $5-30''$ in angular distance on the sky ($1.5-6 \kpc$), if a star formation timescale of $10 \Myr$ (E09) is assumed. This range is measurable with the spatial resolutions of the data. On the other hand, if the timescale is $1 \Myr$ as suggested by T08, the spatial resolution required to see offsets would be $1-5"$.
Therefore, the resolutions of some of the adopted data would be marginal (i.e., 4" in CO, 6" in HI, 1" in H$\alpha$, and 6" in $24\micron$) if $\Delta t$ is as small as found by T08.

In order to test for a possible resolution dependence, we smooth the H$\alpha$ data to match the resolution of the 24$\micron$ data ($\sim$6") and re-measure the offsets, and find practically no appreciable difference
from the measurements at a higher resolution for HI$\rightarrow$H$\alpha$ and CO$\rightarrow$H$\alpha$.
A visual comparison of the H$\alpha_{smoothed}$ maps with the 24$\micron$ maps shows that the two emissions are very well aligned and trace similar emission sources. The difference in resolution of the SF tracers does not appear to be effecting the measured offset.

We also smooth the CO data to the resolution of the HI data (6") and re-measure the offsets with the SF tracers. Again, we find no significant difference in these measurements with the higher resolution images for CO$\rightarrow$24$\micron$ and CO$\rightarrow$H$\alpha$. Therefore, the resolution dependence is negligible for the data that we analyzed.

\subsubsection{Bin Size}\label{sec:binsize}

We also test two different radial bin sizes (5'' and 2'' bins) in measuring offsets. These are the bin sizes used in T08 and E09, respectively. We find little difference between the offset patterns for all four combinations of the tracers; and hence, this is not the cause of the discrepancy.

\section{Discussion}

\subsection{Nature of Gas and Star Formation Tracers}\label{sec:cause}

We found that the most significant cause of the discrepancy between the previous studies is the difference in the gas and star formation tracers used. We also found that CO emission is the best tracer of the compressed gas in spiral arms and that H$\alpha$ is the best tracer of the associated star forming regions. Thus, they are the most useful for the offset measurement among currently available archival data. The previous studies adopted different tracers for tenable and practical reasons. Here we discuss the advantages and disadvantages of each tracer.

H$\alpha$ emission is from HII regions around young massive stars whose lifetimes are short ($\lesssim 10$Myr). There is no doubt that discrete H$\alpha$ peaks pinpoint the locations of very recent star forming regions. In addition, many sets of H$\alpha$ data are readily available in archives, and typically have a higher resolution than the Spitzer $24\micron$ data, another widely-used tracer of star formation.
On the other hand, H$\alpha$ emission could be easily obscured, especially in dense star-forming gas in spiral arms. We might be preferentially finding HII regions far away from the parental gas, biasing the measured offsets toward the large side (T08).
This, however, has proved not to be the case.
The ability of H$\alpha$ emission to trace the locations of star forming regions has been shown recently by a comparison of  H$\alpha$ and Paschen $\alpha$ images of M51 \citep{egusa11}. 
Even in the densest regions in a spiral arm, the majority of HII regions, if not all, can be seen in H$\alpha$ emission, though their fluxes could be significantly attenuated.
H$\alpha$ emission is the best locater of HII regions for offset measurement since their positions are the only parameter required.

The $24\micron$ emission is less sensitive to extinction and might be a better locator of star forming regions that are embedded deeply in a dense star-forming spiral arm.
However, it may not be as straightforward a star formation tracer as previously thought, since the emission could be from dust heated by longer-lived, older stellar populations \citep{liu11}. In addition, the dust could be heated by the continuum emission longward of the Lyman limit from young stars. The shorter-wavelength emission is mostly absorbed in the vicinity of the young stars and forms HII regions, but the longer-wavelength photons can potentially travel farther and heat up dust not immediately associated with the young stars. If this is the case, the $24\micron$ may not pinpoint sites of recent star formation.
The $24\micron$ image is also sensitive to the distributions of both heating sources and the gas/dust.
If the gas spiral arm and recently-formed stellar spiral arm are spatially offset \citep[e.g., the stellar arm should be downstream within a corotation radius;][]{rob69}, the young stars may illuminate the gas/dust spiral arm from the front-side and shift the peak of the $24\micron$ emission.
There is little doubt that young stars are the dominant source of dust heating around star forming regions, and their discrete appearance is very clear in the $24\micron$ image. However, contamination by the escaped photons from HII regions and the older stellar population
could be possibly biasing the offset measurements using 24$\micron$ emission to the small side.

A concentration of CO ($J=1-0$) emission is the clearest tracer of dense molecular spiral arms and is aligned perfectly with narrow dust lanes in optical images \citep{koda09}. The critical density for collisional excitation of the low-$J$ CO line is high \citep[e.g., $\sim \mbox{a few} \times 100 \rm \, cm^{-3}$][]{scoville87}, and all young star forming regions in the Milky Way are associated with giant molecular clouds (GMCs), which are bright in CO. There is no doubt that high-resolution CO data show the locations of the gas spiral arms that are being compressed by the spiral density-wave.

It also had seemed reasonable to assume that the HI emission is adequate to locate the enhancements of gas density in a spiral arm (T08 and F11). However, the small, and often zero, offsets of SF tracers from HI, as opposed to the large offsets from CO, are a clear sign that the HI emission peaks are not tracing the compressed gas in the spiral arm. In fact, the HI peaks are almost always at the downstream side of the CO peaks (Figure \ref{fig:spiralarm} and Figure \ref{fig:cohicompare}). The zero offsets to SF tracers most likely indicate that the HI emission is tracing gas that is photo-dissociated by recent star formation \citep{allen02}. \citet{blitz07} showed that GMCs are often found at the peaks of HI emission in the HI-dominated galaxy M33, but there is also more extended HI emission across the galaxy in regions without GMCs. HI peaks may trace locations where the gas will not form into stars.
It is remarkable that the fraction of molecular gas does not change azimuthally across spiral arms in M51 \citep{koda09}, and therefore the effect of photodissociation is small in terms of the global evolution of the gas phase in the spiral galaxy. However, this small change makes a huge impact in the identification of the HI arm -- since overall the abundance of atomic gas is low -- and likely caused the discrepancy in the previous studies.
Perhaps there is some enhancement of HI emission along the gas spiral arm, but in most cases it is washed out by the stronger emission of the photo-dissociated HI gas, as indicated by the zero/small offsets between HI and SF (T08 and F11).
These observational results require a reconsideration of the notion of a short GMC life-cycle, an argument largely based on the offset measurements with HI and 24$\micron$ emissions (T08). The offset measurement does not indicate the lifetime of GMCs, since the gas stays mostly molecular even after spiral arm passages and star formation.

Therefore, we conclude that CO and H$\alpha$ are the best tracers of the compressed gas in spiral arms and the associated star forming regions, respectively. The HI emission does not necessarily trace the compressed, star-forming gas in the spiral arm; instead it traces predominantly the gas photo-dissociated by recent SF. Emission from 24$\micron$ could be tracing an older stellar population in addition to areas of recent SF and many not pinpoint the site of SF due to the offsets between the dust and young star distributions. 

In this work our discussion of the offset is limited to M51, and we can not make any general conclusion. Nevertheless, our finding that HI traces the photo-dissociated gas, rather than star-forming dense gas, offers a natural explanation for the general discrepancies in the previous studies. The best combination of tracers (CO and H$\alpha$) shows clear spatial offsets and indicates larger, mostly positive offsets than previously suggested using the HI data (T08), although the positive offsets are not ordered as predicted by the standard density-wave theory.

\subsection{Model Limitations to the Offset Measurement}\label{sec:limitations}

The offset method is based on only a few assumptions, i e., a constant $\Omega_{\rm p}$ and $\Delta t$ and pure circular rotation.
Its simplicity assures robustness, but has some weaknesses as well.
We note and summarize several limitations here which are likely related to the large scatter found in the plots presented above.
Some of the limitations come from the model (Equation 1), and the others are from
intrinsic location-to-location variations of gas conditions and star formation in spiral arms.

An intrinsic spatial variation of CO and H$\alpha$ distributions along spiral arms is a source of a systematic error.
As discussed in \S \ref{sec:results}, the spiral arms are not simple, continuous structures; at small scales they appear as ensembles of more discrete clumps in both CO and H$\alpha$ images. The spiral arms go back and forth between the downstream and upstream sides, and this wiggling directly affects the offset measurement. The typical amplitude of the wiggle is about 200 pc and contributes to the scatter in Figure \ref{fig:offsets}. We should note that this type of error tends to cancel out and does not bias offset values systematically. In addition, the molecular gas shows filamentary structures (or spurs) in the interarm regions at the downstream side of spiral arms \citep{koda09}. These interarm structures can potentially limit the offset measurement, especially using the cross-correlation method where such individual structures are neither identified nor rejected in the analysis.


One of the other limitations is the finite lifetime of HII regions. The star formation timescale $\Delta t$ derived by the previous studies ranges between $\lesssim 3 \Myr$ (T08) to $\gtrsim 10 \Myr$ (E09). 

Some of these timescales are as short as (or less than) the typical lifetime of HII regions ($\sim 10\Myr$). The derived timescales are meaningful only when they are comparable with or longer than the the lifetime of HII regions. If one obtains a shorter timescale, more careful consideration, such as taking into account the age distribution of HII regions, is necessary before any physical interpretation of the results.

The assumption of a constant pattern speed is commonly adopted, but it is possible that this assumption is not valid. Indeed, it has been suggested that $\Omega_{\rm p}$ changes radially in M51 \citep{meidtm51}. In particular, the grand-design spiral arms in M51 could be driven by the companion galaxy (NGC 5195). \citet{oh08} discussed, based on theoretical modeling, that $\Omega_{p}$ in tidally-driven spiral arms may vary with radius. The mostly positive offsets that we found in M51 are an indication that the material is flowing through the spiral pattern. Therefore, the spiral needs to be a wave, not a concentration of material. This, however, does not mean that the spiral pattern speed is constant with radius (i.e., the traditional density wave). If the stellar spiral density enhancement lives longer than the gas crossing time, the observed results can be explained without a stationary spiral pattern. Note that the assumption of constant $\Omega_{\rm p}$ has an impact neither on the offset measurement itself nor on our conclusion that the cause of the discrepancy in the previous studies is the choice of tracers.

The circular rotation of galaxies is another assumption that most studies adopt.
T08 calculated the directional changes in gas orbits due to the spiral arm potential
and concluded that its impact on $\Delta t$ is very small (i.e., substantially less than a factor of two).
Another important factor is the change of the amplitude, not only the direction, of gas velocity
due to a symmetric galactic potential.
Gas orbits in spiral galaxies are often described as ovals \citep{wad94, ono04, kod06}.
In an axisymmetric galactic potential the gas moves inward and outward;
the speed being a maximum at the points on the semi-minor axis of the orbit, and a minimum on the semi-major axis.
Spiral arms form primarily due to this slowdown on the semi-major axis; the gas stays longer on the spiral arms,
which leads to the density enhancement. More precisely, the spiral density enhancement occurs even if we do
not account for the shock (which produces even more enhancement), and is proportional to the passage
times through the spiral structure $t_{\rm cross} \propto 1/(\Omega(r)-\Omega_{p})$.
$\Omega(r)$, and therefore $t_{\rm cross}$, change substantially during the orbital motion,
and can change the surface density in spiral arms by an order of magnitude \citep[][ their figure 11]{ono04}.

\begin{figure}
\includegraphics[scale=.75]{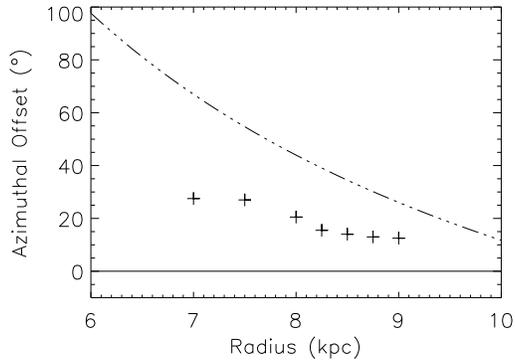}
\caption{
Comparison of offsets from two models: the analytical circular rotation model adopted
in most offset studies (dash-dotted line) and more realistic numerical simulations
using a stationary spiral pattern by \citet{dobb10} and F11 (crosses).
The simulations include non-circular motions across spiral arms,
and therefore trace any slowdown of gas and stellar motions.
The two models adopt the same global pattern speed, $\Omega_{p}$ = 20 $\kmps$ kpc$^{-1}$,
and a flat rotation curve with a peak velocity of 220 kms$^{-1}$.
The offsets are measured between the gas and a 100 Myr-old stellar population.
The offsets in the numerical simulations are lower by a factor of $\sim 2$
than those predicted by the circular rotation model, indicating that
the gas and stars slow down considerably during spiral arm passage.
}
\label{fig:100myr}
\end{figure}

To estimate the effect of this slow down to $\Delta t$, we estimate a difference in rotational velocities between circular orbits assumed in Eq. (1) and elliptical orbits that form spiral arms.Assuming that the gas is traveling in an elliptical orbit with the eccentricity $e=\sqrt{1-b^{2}/a^{2}}$ (maximum and minimum distances $a$ and $b$, respectively) and that the galactic potential is isothermal (i.e., flat rotation curve, $\Phi \propto ln(r)$),
conservation of energy and angular momentum provide
\begin{equation}
\frac{v_{a}}{v_{c}}= \frac{\Omega_{a}}{\Omega_{c}}=\sqrt{\ln \left(1-e^{2}\right)\frac{\left(e^{2}-1\right)}{e^2}}
\end{equation}
where $v_{a}$ is the rotation velocity at the maximum distance $a$ for an elliptical orbit, $v_{c}$ is the circular rotation velocity at radius $a$. $\Omega_{a}$ and $\Omega_{c}$ are the angular speeds corresponding to $v_a$ and $v_c$.
For example, assuming an elongated orbit ($e=0.6$-$0.8$), derived pattern speed ($\Omega_{\rm p}=30 \kmps \kpc^{-1}$), and rotation velocity ($\sim 200\kmps$), $\Delta t \propto 1/(\Omega-\Omega_{\rm p})$ changes from its value under circular rotation by a factor of 2-3 (but up to 10-30) at radii of 4-6 kpc .
If the non-circular motion, the very essence of spiral arms, is taken into account, the derived $\Delta t$ under the assumption of circular rotation is an underestimate by a factor of 2-3.
It is important to note that the non-circular motions do not explain the discrepancy in the measured offsets. They affect the interpretation of the azimuthal offsets, but not their measurements. [Note again that the main focus of this study is the measurements.] In order to confirm our simple analytic calculations, we compare the simplistic circular rotation model with the numerical simulations by \citet{dobb10} and F11 that include non-circular motion. We use the offsets measured by F11 between  the gas and 100 Myr old clusters in the simulations. The pattern speed of their stationary spiral model is 20 $\kmps$ kpc$^{-1}$ and the rotation curve is roughly flat with a peak velocity of 220 $\kmps$. We adopt these parameters for our circular rotation model. Figure \ref{fig:100myr} shows the expected offsets for the circular rotation model, Eq. (1), compared to the offset from the numerical simulations. It is clear that the circular rotation model overpredicts the offsets by a factor of two in this radius range. Therefore, the gas motion across the spiral arm slows down in the more realistic numerical simulations. The difference is expected to be larger around the co-rotation radius ($\sim$ 11 kpc in this parameter set). These results are consistent with our simple calculation of non-circular orbits. We conclude that the circular rotation model adopted in the previous studies tends to underestimate the SF timescale by a factor of 2-3.
Development of an orbit model that fits M51 is beyond the scope of this paper,
but this discussion provides a caveat that the $\Delta t$ derived from the offset method is likely a lower limit of the SF timescale,
which could be about an order of magnitude greater when the non-circular motion is taken into account. This could also be evidence against a short GMC lifetime.

\section{Conclusion}
The discrepancy found between the previous offset measurements (T08, E09, F11) is significant. E09 found relatively large offsets, which are consistent with the predictions of the density-wave theory and possibly gravitational collapse of the dense gas formed in spiral arms. T08 found offsets an order of magnitude smaller, which would suggest extremely rapid star formation and destruction of giant molecular clouds in a spiral arms (i.e., faster than the free-fall timescale of the gas). Most striking are the non-ordered offsets found by F11, which could argue against the traditional density-wave theory. To elucidate the cause of the discrepancy, we repeated the previous studies using the gas tracer emissions (CO and HI) and SF tracer emissions (H$\alpha$ and 24$\micron$) and applying the two measurement methods that were developed in the previous studies.

We analyzed the spiral arms in M51 and found that the primary cause of the discrepancy is the use of different gas tracers. In particular, the HI 21 cm line emission traces predominantly the gas photo-dissociated by recently-formed stars, but not necessarily the compressed, star-forming gas in spiral arms. In fact, the HI peaks are almost always at the downstream side of the CO peaks. This explains the small or non-ordered offsets between HI and SF tracers found by T08 and F11 using HI data. It is important to use CO emission to trace the parental gas for SF. In our comparison of CO and H$\alpha$ emission we found mostly positive spatial offsets with substantial scatter. The positive offsets suggest that there is a density wave and that material is flowing through the spiral arm. However, this may not be a density wave of the simplest form with a single pattern speed. The ability of H$\alpha$ emission to locate the positions of young star-forming regions (even in very dense environments) has been confirmed previously by \citet{egusa11}, although its flux suffers significantly from large dust attenuation.

We compared several other differences in the analyses of the previous studies, and found that nothing but the tracers contribute significantly to the discrepancy. The differences that we compared include those in the offset measurement methods, spatial resolution, and bin size. The dominant cause of the discrepancy is the different choice of the tracers in the previous studies. HI emission does not necessarily trace the sites of compressed gas nor star formation, and contaminated significantly by photodissocation due to young stars.


We thank Jim Barrett for useful suggestions. We would also like to thank the anonymous referee for comments that have improved this paper. This work is supported in part by the NSF under grant AST-1211680. JK also acknowledges support from NASA through grant NNX09AF40G, an Hubble Space Telescope grant, and a Herschel Space Observatory grant.





\begin{deluxetable}{ccccccc}
\tablecolumns{7}
\tablewidth{0pc}
\tablecaption{Summary of Previous Offset Studies}
\tablehead{
\colhead{} & \multicolumn{2}{c}{Tracers} & \colhead{Analysis} & \colhead{Method} & \colhead{$t_{\rm SF}$} & \colhead{$\Omega_{\rm p}$} \\
\cline{2-3} \\
\colhead{} & \colhead{Gas} & \colhead{SF} & \colhead{} & \colhead{} &\colhead{(Myr)} & \colhead{($\kmps \,\rm kpc^{-1}$)}}
\startdata
Tamburro et al. 2008 & HI & $24\micron$ & Arm 1\&2 & CC & $3.4 \pm 0.8$ & $21 \pm 4$ \\
Egusa et al. 2009 & CO & H$\alpha$ & Arm 1\&2 & PT & 13.8 $\pm$ 0.7 & 40 $\pm$ 4 \\
Egusa et al. 2009 & CO & H$\alpha$ & Arm 1 & PT & 7.1 $\pm$ 0.5 & 31 $\pm$ 5 \\
Foyle et al. 2011 &  HI &  24 \micron & Arm 1\&2 & CC & \multicolumn{2}{c}{No ordered offsets}  \\
\enddata
\label{tab:previouswork}
\end{deluxetable}


\end{document}